\documentclass{article}
\usepackage{subcaption}
\usepackage{graphicx} % Required for inserting images
\usepackage{amsmath}
\usepackage[noadjust]{cite}
\usepackage{hyperref}
\usepackage{cleveref}

\usepackage[a4paper, total={6in,8in}]{geometry}
\usepackage{graphics}
\usepackage{xcolor}

\usepackage[utf8]{inputenc}
\usepackage[T1]{fontenc}
\usepackage{amsmath,amssymb,amsthm}
\usepackage{easybmat}
\graphicspath{ {./images/} }
\usepackage{blindtext}

\title{Performance Analysis of Satellite-Based QKD Protocols}
    \author{
Muskan\thanks{E-mail: muskan.1@iitj.ac.in},
Ramniwas Meena\thanks{E-mail: meena.53@iitj.ac.in},
Subhashish Banerjee\thanks{E-mail: subhashish@iitj.ac.in}
\\
Department of Physics, Indian Institute of Technology, Jodhpur, India-342030\\
}
\date{December 2025}

\begin{document}

\maketitle
\begin{abstract}
Satellite-based free-space quantum key distribution (QKD) provides a practical framework for achieving secure global communication beyond the limitations of optical fibers. In this work, the quantum bit error rate (QBER) and secure key rate of four representative protocols-BB84, B92, BBM92, and E91 are investigated over low earth orbit (LEO) links in both uplink and downlink configurations. The optical link is modeled using a Gaussian beam formalism, incorporating the effects of diffraction, pointing errors, atmospheric turbulence, and background noise contributions. The protocols are examined under day and night-time operating conditions, and their dependence on the zenith angle is analyzed. The findings show that downlink links generally exhibit lower QBER and higher secure key rates than uplinks, and among prepare-and-measure schemes, BB84 consistently outperforms B92, while in entanglement-based approaches, BBM92 achieves higher key rates than E91. 
\end{abstract}

\providecommand{\keywords}[1]
{
  \small  
  \textbf{\textit{Keywords---}} #1
}

\section{Introduction}
Quantum communication, particularly Quantum Key Distribution (QKD), offers a robust method for secure data transfer, with quantum identity authentication ensuring that only legitimate users are verified~\cite{CS01, dutta2022short}. In such systems, photons act as the primary information carriers, enabling high transmission rates in optical fiber infrastructures. Nevertheless, the extension of quantum links over large distances is fundamentally limited by photon attenuation during propagation~\cite{MPR+19}. While quantum repeaters have been proposed to counteract this loss~\cite{ATL15, ZPD+18, SGL18}, their deployment remains technically demanding and is not yet practical. At present, satellite-assisted free-space channels represent the most promising solution for achieving global-scale QKD~\cite{P3021}. These links benefit from advances in satellite engineering and optical communication, alongside strategies developed to mitigate the effects of noise in quantum channels~\cite{SB18}. Both theoretical investigations~\cite{ELH+21, CSH+22} and experimental demonstrations confirm that satellite-based QKD has reached a stage of technological readiness for real-world implementation. A crucial tool in designing such systems is asymptotic key rate analysis, which enables optimization of parameters to maximize secure key generation~\cite{ABN21, dutta2024analysis, 2025finite, dutta2025satellite}. Recent studies have also extended this approach to continuous-variable \cite{meena2025continuous, KGW+24} and discrete-variable QKD \cite{KM25} in scenarios where the eavesdropper’s access is restricted via bypass channels~\cite{GBL+23}. This work has shown that, under high-loss conditions, limiting an adversary’s information pathway can significantly enhance achievable key rates.
\par

The first QKD protocol, known as BB84, was introduced by Bennett and Brassard in 1984~\cite{bennett2020quantum}. Since then, both theoretical studies and practical implementations of QKD have progressed steadily~\cite{scarani2009security, shenoy2017quantum, pirandola2020advances}. Despite this progress, the practical deployment of QKD is still limited by the difficulties in generating, maintaining, and manipulating quantum resources with current technology. This has encouraged the development of protocols with simpler designs that require fewer quantum resources. For example, the BB84 protocol uses four different quantum states and two measurement bases. In 1992, Bennett proposed a simpler protocol, called B92, which uses only two nonorthogonal states and two measurement bases~\cite{PhysRevLett.68.3121}. However, as noted in the original work~\cite{PhysRevLett.68.3121}, the B92 protocol is more sensitive to noise compared to BB84 and other QKD protocols. The E91 protocol, introduced by Ekert in 1991, employs entangled photon pairs and verifies security through the violation of Bell’s inequalities~\cite{ekert1991quantum}. While this approach provides a strong, device-independent security foundation, it requires multiple measurement settings and dedicates a portion of the detected events to Bell tests rather than key generation, reducing overall efficiency. To simplify implementation and improve key generation rates, Bennett, Brassard, and Mermin proposed the BBM92 protocol in 1992~\cite{bennett1992quantum}, which retains the use of entangled states but adopts the measurement and sifting procedure of BB84. This modification eliminates the need for Bell inequality testing, reduces experimental complexity, and allows a larger fraction of detected events to contribute to the sifted key, making BBM92 more practical for high-rate quantum key distribution.\par
The relevance of such comparisons becomes even more pronounced in satellite-based QKD, where physical channel characteristics introduce asymmetric challenges. Depending on the transmission direction, links can be classified into two configurations: uplink and downlink. These configurations are not symmetrical, as the sequence in which the signal beam traverses the atmosphere and outer space differs significantly. In the uplink case, the beam first propagates through the atmosphere, where it is subject to turbulence and scattering, before continuing through the vacuum of space, where beam broadening becomes the dominant effect. In contrast, for downlink transmission, the beam initially travels through space and subsequently passes through the atmosphere, with pointing errors being the primary factor influencing its propagation over long distances. This difference in traversal order imposes distinct technical requirements on ground-based and spaceborne receiving systems~\cite{bourgoin2013comprehensive}.\par
This work investigates the performance of four QKD protocols-BB84, B92, BBM92, and E91-representing both prepare-and-measure and entanglement-based schemes \cite{SOV+24}, for the computation of asymptotic key rates in satellite-based QKD using guassian beam distribution ~\cite{bourgoin2013comprehensive}. The analysis is carried out for both uplink and downlink configurations, incorporating atmospheric transmittance values obtained from MODTRAN~6 \cite{berk2014modtran}. Background noise, including stray counts, is considered for both transmission directions under day and night conditions for a low Earth orbit (LEO) satellite at an altitude of 500~km. Compared to medium Earth orbit (MEO) and geostationary Earth orbit (GEO) satellites, the shorter propagation distance in LEO reduces free-space losses, increases photon detection probability, and thereby enhances secure key generation rates. Beam propagation is modeled using a Gaussian beam formalism~\cite{bourgoin2013comprehensive}, providing a realistic description of optical transmission in LEO-based QKD links. The model incorporates beam broadening due to diffraction and Gaussian-distributed pointing errors, while for uplink scenarios it additionally accounts for atmospheric turbulence effects, characterized using the Hufnagel-Valley turbulence profile~\cite{valley1980isoplanatic, hufnagel1964modulation}. By incorporating atmospheric modeling, orbital configuration, and background noise effects into a unified framework, this work demonstrates how the resource requirements of different QKD protocols interact with the inherent asymmetries of uplink and downlink satellite channels. The findings provide a clear comparison of the relative strengths and limitations of BB84, B92, BBM92, and E91 in realistic low-Earth orbit scenarios, offering practical guidelines for protocol selection in future large-scale QKD networks.
The remainder of this paper is organized as follows. Section~\ref{QKD Protocols} provides an overview of the BB84, B92, BBM92, and E91 protocols. Section~\ref{Channel Modeling} describes the channel modeling framework, including the relevant loss mechanisms. Section~\ref{Environmental Photons} discusses environmental noise with a focus on stray photon counts. In Section~\ref{QBER}, the QBER is analyzed for the considered protocols, while Section~\ref{Keyrate} presents the corresponding key rate formulations. Section~\ref{Numerical Results and Discussion} reports the numerical results and offers a detailed discussion. The paper concludes with a summary of the key points.

\section{QKD Protocols}\label{QKD Protocols}
QKD protocols are the foundation of secure communication enabled by quantum mechanics. These protocols establish a shared secret key between two parties, typically referred to as Alice and Bob, while ensuring that any eavesdropping attempts are detectable. Several QKD protocols have been developed, each with its unique approach to key generation and security guarantees \cite{TTN+25, AWB+25}. Here we briefly discuss the QKD protocols used in this work:
\subsection{BB84}

The BB84 protocol, which was proposed in \cite{bennett2020quantum}, is a QKD protocol. It aims to facilitate the secure generation and sharing of a secret cryptographic key between two parties, typically referred to as Alice and Bob, for subsequent communication purposes. In this protocol, Alice prepares a random string of qubits, which can be in the states $ |0\rangle$, $ |1\rangle$, $ |+\rangle$, or $ |-\rangle$, and transmits them to Bob. Upon receiving the qubits, Bob randomly measures each qubit in either the standard basis ${0,1}$ or in the Hadamard basis  ${+,-}$  and publicly announces the basis used for each measurement. Alice then informs Bob about the cases where Bob's basis matches her own. Only the qubits measured using the same basis by both Alice and Bob are retained, while the rest are discarded. It is expected that, on average, Bob's basis will match Alice's basis $50\%$ of the time. In an ideal scenario, the retained qubit strings of Alice and Bob are perfectly symmetric. To verify the correctness of the generated key, Bob uses a portion of the string as a verification string and publicly announces the measurement results and corresponding positions of these qubits. Alice compares Bob's results with her own. If the number of errors exceeds a tolerable limit, the protocol is discarded. Otherwise, a perfectly symmetric, random, and unconditionally secure quantum key is obtained.

\subsection{B92}
  The B92 protocol, a simplified variant of the BB84 protocol, is another QKD protocol \cite{PhysRevLett.68.3121}. In contrast to BB84, the B92 protocol utilizes only two quantum states, denoted as $|0\rangle$ and $ |+\rangle$, for information transmission. In this protocol, Alice randomly sends a sequence of qubits in states $|0\rangle$ and $|+\rangle$ to Bob, where $ |0\rangle$ represents bit value $0$, and $ |+\rangle$ represents bit value $1$. The primary distinction between BB84 and B92 lies in the number of states employed for transmission. BB84 employs four states ($ |0\rangle$, $ |1\rangle$, $ |+\rangle$, $|-\rangle$), while B92 employs only two ($|0\rangle$, $ |+\rangle$). Bob performs measurements randomly in one of two bases: {$ |0\rangle$, $ |1\rangle$} or {$ |+\rangle$, $|-\rangle$}. However, in the B92 protocol, Bob does not disclose the basis used for measurement; instead, he retains qubits where the measurement outcome is $|1\rangle$ or $|-\rangle$, discarding the others. Bob later reveals which qubits he retained. Following Bob's disclosure, Alice retains the corresponding qubits (partner qubits) and discards the remainder. If Bob measures $|0\rangle$, he cannot determine whether Alice sent $|0\rangle$ or $|+\rangle$ because the $ |+\rangle$ state measured in the computational basis collapses to $ |0\rangle$ half the time. Thus, if Bob's measurement yields $|0\rangle$, he cannot deduce the encoding used by Alice. Moreover, if Alice sends $|0\rangle$, Bob will never receive it as $|1\rangle$. This is because if Bob selects the computational basis, he will always measure it as $|0\rangle$, and if he chooses the diagonal basis, he will obtain $ |+\rangle$ or $ |-\rangle$ with equal probability. Hence, Bob can only obtain $|1\rangle$ if Alice sent $|+\rangle$. Consequently, when Bob's measurement yields $|-\rangle$, he can infer that Alice sent $ |0\rangle$. However, he cannot make any conclusions when the measurement yields $ |+\rangle$. As Bob is aware of the encoding, the remaining string can be utilized to generate a random symmetric key.

\subsection{E91}
The E91 protocol is an entanglement-based QKD scheme \cite{ekert1991quantum} in which a source emits pairs of maximally entangled spin-\(\tfrac{1}{2}\) particles (or polarization-entangled photons) in the singlet state. One particle from each pair is sent to Alice and the other to Bob. For each received particle, Alice and Bob independently and randomly choose one of three analyzer orientations lying in the transverse (\(x\!-\!y\)) plane. The azimuthal angles defining Alice’s settings are
\[
\phi_1^{(a)}=0,\quad \phi_2^{(a)}=\tfrac{\pi}{4},\quad \phi_3^{(a)}=\tfrac{\pi}{2},
\qquad
\phi_1^{(b)}=\tfrac{\pi}{4},\quad \phi_2^{(b)}=\tfrac{\pi}{2},\quad \phi_3^{(b)}=\tfrac{3\pi}{4}.
\]
Each local measurement yields a dichotomic outcome \(x,y\in\{+1,-1\}\). When the two parties happen to select the same orientation, the singlet correlations produce perfect anticorrelation, enabling those outcomes to be mapped to raw key bits after sifting.

Outcomes obtained with \emph{different} analyzer settings are used to estimate nonlocal correlations. Let \(P_{\alpha\beta}(a_i,b_j)\) denote the joint probability of outcomes \(\alpha,\beta\in\{+1,-1\}\) given settings \(a_i\) and \(b_j\). The correlation function is
\begin{equation}
E(a_i,b_j)= P_{++}(a_i,b_j)+P_{--}(a_i,b_j)-P_{+-}(a_i,b_j)-P_{-+}(a_i,b_j),
\label{eq:Eij}
\end{equation}
which, for the singlet, is equivalent to the quantum prediction \(E(a_i,b_j)=-\mathbf{a}_i\!\cdot\!\mathbf{b}_j\).
To certify security, Alice and Bob evaluate the CHSH parameter using a subset of setting pairs:
\begin{equation}
S=E(a_1,b_1)-E(a_1,b_3)+E(a_3,b_1)+E(a_3,b_3).
\label{eq:CHSH}
\end{equation}
Local-hidden-variable models satisfy \(|S|\le 2\), whereas quantum mechanics requires \(|S|=2\sqrt{2}\) for appropriately chosen angles. Any eavesdropping attempt that gains information necessarily alters the two-party statistics and diminishes the Bell violation, driving the observed \(S\) toward the classical bound.

After parameter estimation (Bell test) confirms a statistically significant violation of Eq. \eqref{eq:CHSH}, the parties keep only the measurement results obtained with identical analyzer orientations to form the sifted key. In this way, E91 integrates key generation with an inherent security test grounded in Bell’s theorem.

\subsection{BBM92}
The BBM92 protocol, introduced by Bennett, Brassard, and Mermin, is an another entanglement-based quantum key distribution scheme \cite{bennett1992quantum} that avoids reliance on Bell's theorem. In this protocol, an entangled source distributes photon pairs to Alice and Bob, who each randomly choose to measure along the $Z$ or $X$ basis. After measurement, they announce only their basis choices and retain the data corresponding to matching bases, which are perfectly correlated (or anti-correlated) and thus form the raw key. To detect eavesdropping, a subset of outcomes is publicly compared; any adversarial disturbance would break the correlations and produce detectable errors. Unlike Ekert's E91 protocol, BBM92 does not require testing Bell inequalities—its security follows directly from the impossibility of faking entanglement correlations without introducing observable errors. Conceptually, BBM92 can be seen as the entanglement-based counterpart of BB84: in BB84, Alice actively prepares and sends single-qubit states, while in BBM92, the randomness arises naturally from entanglement measurements. This intrinsic symmetry makes BBM92 a foundational protocol for entanglement-based QKD systems and a bridge between prepare-and-measure and entanglement-assisted approaches.

\section{Channel Modeling}\label{Channel Modeling}
 To assess the practicality of satellite QKD, a numerical framework was employed to estimate the overall channel loss. The loss from the diffraction is calculated by the Rayleigh–Sommerfeld diffraction \cite{Goodman1996} . After considering the loss, the intensity at the receiver located at position $\vec{l}$,

\begin{equation}
I_1(\vec{l}) =  \left| \frac{d^2}{\lambda^2}\iint_{S_t} \sqrt{I_0(\vec{l}\,')} \cdot \frac{1}{|\vec{l} - \vec{l}\,'|^2} \cdot \exp\left( \frac{2\pi i}{\lambda} |\vec{l} - \vec{l}\,'| \right) \, dx\,dy \right|^2,
\end{equation}
where $\vec{l'}$ represents a point on the transmitter surface $S_t$, over which the integration is performed. $I_0$ is the intensity distribution at the transmitter, $\lambda$ is the wavelength of the optical beam, and $d$ is the propagation distance between the satellite and ground station. Since the beam exhibits circular symmetry, the intensity can be evaluated along $x=0$ (or $y=0$), from which the radial distribution $I(r)$ is determined, where $r$ denotes the distance from the beam center.
 \\
The loss resulting from pointing errors is analyzed by examining the distribution of the beam center at the receiver. This distribution is modeled as a two-dimensional Gaussian function representing the pointing error, expressed as:
\begin{equation}
g(r) = \frac{1}{2\pi\sigma_p^2} \exp\left(-\frac{r^2}{2\sigma_p^2}\right).
\end{equation}
Here, $\sigma_p$ denotes the standard deviation associated with the pointing error. The resulting beam intensity profile at the receiver, denoted by $I_2(\vec{l})$, is obtained by performing a two-dimensional convolution \cite{baddour2009operational} of the diffracted beam $I_1$ with the Gaussian distribution $g$, which accounts for the pointing error at the receiver. This can be expressed as:
\begin{equation}
I_2(\vec{l}) = (I_1 * g)(r,\theta) = \int_0^{2\pi} d\theta' \int_0^{\infty} I_1(r')\, g(r - r')\, dr'.
\end{equation}
For uplink transmission, it is important to account for the beam broadening induced by atmospheric turbulence, which evolves over a characteristic timescale of the order of 10-100~ms. When the beam profile is averaged over durations substantially longer than this timescale, the resulting spatial distribution converges to a Gaussian form. Based on the Hufnagel--Valley model of atmospheric turbulence~\cite{accetta1993infrared} the beam waist $w$ at the receiver is given by:
\begin{equation}
w = \frac{2\sqrt{2}d\lambda}{\pi r_0}
\end{equation}
where $r_0$ is the transverse coherence length, can be expressed as:
\begin{equation}
r_0 = \left[ 1.46 \sec({\phi})\left( \frac{2\pi}{\lambda} \right)^2 \int_0^h C_n^2(z) \left(1 - \frac{z}{h} \right)^{5/3} dz \right]^{-3/5},
\end{equation}
where $\phi$ denotes the zenith angle, $h$ is the altitude of the receiver, and $C_n^2(z)$ represents the refractive-index structure constant as a function of altitude as given by:
\begin{equation}
C_n^2(z) = 0.00594\left(\frac{v}{27}\right)^2(z \times 10^{-5})^{10} e^{-z/1000} + 2.7 \times 10^{-16} e^{-z/1500} + A e^{-z/100}.
\end{equation}
The coefficients $v$ and $A$ characterize the influence of atmospheric turbulence and are defined based on topological atmospheric conditions. For this analysis, we adopt typical nighttime sea-level values of $A = 1.7 \times 10^{-14}~\mathrm{m}^{-2/3}$ and $v = 21~\mathrm{m/s}$~\cite{bourgoin2013comprehensive}. After determining the waist of the Gaussian distribution, the overall beam profile $I_3(\vec{v})$ is obtained by applying a two-dimensional convolution of the beam profile $I_2(\vec{v})$ with the turbulence-induced broadening.
Once the beam intensity distribution at the receiver has been determined-whether for an uplink or downlink scenario (where $I_3 = I_2$); the corresponding received optical power $P$ is calculated by integrating the intensity profile $I_3(\vec{v})$ over the receiver aperture:
\begin{equation}
P = \iint_{S_r} I_3(\vec{v})\, dx\, dy,
\end{equation}
where $S_r$ denotes the surface area of the receiving telescope. The resulting optical power is directly proportional to the mean number of detected photons.
The impact of atmospheric transmission and detector efficiency is accounted for by scaling the received optical power with the detector efficiency, $\eta_d$, and the atmospheric transmittance, $\eta_t$. The values of $\eta_t$ for both uplink and downlink scenarios are obtained from MODTRAN 6 \cite{berk2014modtran} simulations, using the modtran parameters specified in \cite{bourgoin2013comprehensive}. Finally, the ratio of the received power to the transmitted power $P_0$ is expressed in decibels to compute the total channel loss. An additional fixed loss of 3~dB is included to account for imperfections in the polarization analyser and intrinsic losses associated with optical components and telescope coupling \cite{bourgoin2013comprehensive}. 
\begin{equation}
L = -10 \log_{10} \left( \frac{\eta_t \eta_d P}{P_0} \right) + 3. \label{eq:loss}
\end{equation}

\section{Environmental Photons (stray-photons) \label{Environmental Photons}}
In free-space communication, environmental photons represent a significant noise source. Here, we'll summarize the analysis concerning the impact of these environmental photons on the detector, for both the transmission from a satellite to the ground (downlink) and from the ground to a satellite (uplink), which is crucial in calculating the QBER. We solely focus on the scenario of uplinks during nighttime operation. If the ground station site has a low level of light pollution, the biggest fraction of environmental photons comes from the sunlight reflected first by the moon and then by the earth \cite{liorni2019satellite}.
\begin{equation}
    N_{night}^{up} = A_E A_M R_M^2 a^2 \frac{\Omega_{fov}}{d_{EM}^2} B_f \Delta t H_{sun}.
\end{equation}
In this equation, the symbols denote the following: $A_M$ corresponds to the albedo of the Moon, $R_M$ denotes the Moon's radius, $A_E$ represents the Earth's albedo, and $d_{EM}$ signifies the distance between the Earth and the Moon. $H_{sun}$ is an indicator of the solar spectral irradiance measured in photons per second per nanometer per square meter at the specific wavelength under consideration. $\Omega_{fov}$ and $a$ represent the angular field of view (AFOV) and the receiving telescope's radius, respectively. $B_f$ denotes the spectral filtering width, and $\Delta t$ signifies the detection time-window.\\
The assessment of background photons in downlinks varies significantly depending on the location. The formula to express a telescope's receiving power is as follows:
\begin{equation}
    P_b = H_b \Omega_{fov} \pi a^2 B_f.
\end{equation}
As a result of the weather and the hour of the day, the parameter $H_b$ determines the total brightness of the sky background. The number of photons per time window can be calculated from the equation above-
\begin{equation}
    N^{down} = \frac{H_b}{h\nu} \Omega_{fov} \pi a^2 B_f \Delta t, 
\end{equation}
In this equation, the symbol $h$ represents the Planck constant, while the symbol $\nu$ represents the frequency of the background photons, which have been filtered.

\begin{table}
\centering
\begin{tabular}{|c|c|c|}
\hline
 Parameter & Value & Brief description  \\
 \hline
$\lambda$ & $785\,\text{nm}$ & Wavelength of the signal light \\
\hline
$r_T$ & $15\,\text{cm},\,50\,\text{cm}$ & Downlink, Uplink \\
\hline
$a$ & $50\,\text{cm},\,15\,\text{cm}$ & Downlink, Uplink \\
\hline
$\eta_d$ & $0.5$ & Detector efficiency \\
\hline
$p_{dark}$ & $5\times10^{-8}$ & Detector efficiency \\
\hline
$\sigma_P$ & $1.2\times10^{-6}$ & Pointing error \\
\hline
 $H$ & 500 km & Minimum altitude (zenith)  \\
\hline
$H_b$ & $1.5\times10^{-6} W m^{-2} sr^{-1} nm^{-1}$ & Night, clear sky  \\
\hline
 $H_b$ & $1.5\times10^{-3} W m^{-2} sr^{-1} nm^{-1}$ & Day, clear sky  \\
\hline
AFOV, $\Omega_{fov}$ & $(100\times10^{-6})^2 sr$ & Night-time downlink  \\
\hline
AFOV, $\Omega_{fov}$ & $(10\times10^{-6})^2 sr$ &     Day-time downlink  \\
\hline
AFOV, $\Omega_{fov}$ & $(30\times10^{-6})^2 sr$ & Night-time uplink  \\
\hline
 $\Delta t$  & $1 ns$ & Night and day time \\
\hline
$B_f$ & $1 nm$ & Night-time downlink  \\
\hline
 $B_f$ & $0.2 nm$ & Day-time downlink  \\
\hline
 $B_f$ & $1 nm$ & Night-time uplink  \\
\hline
$H_{sun}$ & $4.610\times10^{18}$ $phot$ $s^{-1}$ $nm^{-1}$ $m^{-2}$ & Solar spectral irradiance  \\
\hline
$A_E$ & $0.300$ & Earth's albedo  \\
\hline
$A_M$ & $0.136$ & Moon's albedo  \\
\hline
$R_M$ & $1.737\times10^6 m$ & Moon's radius  \\
\hline
$d_{EM}$ & $3.600\times10^8 m$ & Earth-moon distance  \\

\hline
\end{tabular}
\caption{Parameters related to the atmospheric weather conditions, stray photons and environmental light \cite{liorni2019satellite} }
\label{table}
\end{table}

\section{QBER}\label{QBER}
 The QBER is a measure of the ratio of incorrect bit counts to the total number of received bit counts. It is used to quantify the probability of obtaining a false detection in comparison to the total probability of detection per pulse. The QBER is influenced by three main components: the signal component, the dark count component and the stray count component i.e. environmental photons. 
\subsection{\textbf{ QBER for BB84}}
In BB84 protocol, the QBER can be calculated as \cite{lutkenhaus2000security, liorni2019satellite}:
\begin{equation}
    e_{84} = \frac{c \; p_{signal}+\frac{1}{2} \; (p_{dark}+p_{straycounts})}{p_{click}} .
\end{equation}
Here, $c$ corresponds to the error rate associated with depolarization in the encoding degree of freedom or imperfection of the preparation or detection stage leading to incorrect state discrimination. The depolarization error rate is assumed to be $c = 1\%$  \cite{LCX+24, lutkenhaus2000security}, which is a typical value reported in prior studies , accounting for polarization misalignment and optical imperfections. This parameter contributes linearly to the overall QBER; however, small variations in its value (e.g., in the range $0.5\%$-$2\%$) \cite{liorni2019satellite} do not significantly alter the system performance, as the dominant effects are governed by channel loss and background noise. $p_{click}$ represents the overall anticipated probability that Bob will observe the detection of a photon during a specific pulse. Typically, $p_{click}$ is determined by considering three distinct sources that can independently trigger a detection event. These sources encompass photons transmitted by Alice, background dark counts and straycounts.
\begin{equation}
    p_{click} = p_{signal}+p_{dark}+p_{straycounts}.\label{eq:pclick}
\end{equation}
The probability of Bob's detector firing due to a photon emitted by Alice's source is denoted as $p_{signal}$ and is given by-
\begin{equation}
     p_{signal}=1-\exp(-\eta_d \eta \mu), 
\end{equation}
where $\eta_d$ is the detector efficiency, $\eta$ is the total transmittance efficiency and is calculated using Eq.\eqref{eq:loss} and $\mu$ is the average number of photons per pulse.
On the other hand, $p_{dark}$ represents the probability of a dark count occurring in Bob's detector \cite{lutkenhaus2000security}. $p_{straycounts}$ is the probability of occuring straycounts which can be calculated using the expressions from Sec. \ref{Environmental Photons}. It is important to note that the Eq.\eqref{eq:pclick} assumes the neglect of simultaneous occurrences of signal, dark count and straycounts events when all three $p_{signal}$, $p_{dark}$ and $p_{straycounts}$  are small.\par
\subsection{\textbf{ QBER for B92}}
In the B92 protocol, the logical bits are encoded using two non-orthogonal basis states. As a result, in $50\%$ of the cases the coding and decoding are performed in the same basis, while in the remaining $50\%$ the bases differ. The number of usable bits equals $25\%$ of the total received bits \cite{EtenguAbbouWongAbidNortizaSetharaman}. The QBER for the B92 protocol can be calculated as: 

\begin{equation}
    e_{92} = \frac{c \; p_{signal}+\frac{1}{4} \; (p_{dark}+p_{straycounts})}{p_{click}} .\\  
\end{equation}

Here, the parameters have  the same definitions for $p_{signal}$,  $p_{dark}$, $p_{straycounts}$ and  $p_{click}$  as in the BB84 protocol.

\subsection{\textbf{ QBER for BBM92}}\label{QBER for BBM92}
The QBER for BBM92 depends on various factors such as the properties of the quantum channel, the quality of the detectors used, and the presence of any eavesdroppers \cite{PhysRevA.65.052310}. In one beam splitter with transmission, we combine all losses to each receiver from the channel, detectors, and optics.
\begin{equation}
    \alpha_L = \eta_d*\eta.
\end{equation}
The parameter $\eta_d$ represents the detector efficiency of the system. The $p_{coin}$ that represents the coincidence probability is divided into three components: $p_{true}$, which denotes the probability of a genuine coincidence between a pair of entangled photons, $p_{false}$  which represents the probability of a false coincidence and  $p_{straycounts}$ which represents the probability of straycounts.
\begin{equation}
    p_{coin}= p_{true}+p_{false}+p_{straycounts}.
\end{equation}
We must choose a location for the source. Setting the source at a distance of $(L-x)$ from Bob and $x$ from Alice, we get
\begin{equation}
    p_{true}= \alpha_x\alpha_{L-x} =\eta_d\alpha_L.
\end{equation}
\begin{equation}
    p_{false}= 4\alpha_x p_{dark}+4\alpha_{L-x}p_{dark}+ 16p_{dark}^2.
\end{equation}
Keeping only terms which are second order in $\alpha_x$ and $p_{dark}$, it can be observed that the probability of a true coincidence remains constant with respect to $x$, while the false coincidence rate changes. By performing a straightforward optimization, it can be determined that the false coincidence rate reaches its minimum value at a distance halfway between Alice and Bob. The value of this minimum false coincidence rate can be calculated using the given formula:
\begin{equation}
    p_{false}=8\alpha_{L/2}\; p_{dark} + 16\; p_{dark}^2 .   
\end{equation}
The QBER is given by \cite{PhysRevA.65.052310}
\begin{equation}
    e_{M92} = \frac{c \; p_{true}+\frac{1}{2} \; (p_{false}+p_{straycounts})}{p_{coin}}.
\end{equation}
\subsection{\textbf{ QBER for E91}}
QBER for $E91$ is given by \cite{ekert1991quantum, PhysRevA.65.052310}
\begin{equation}
     e_{91} = \frac{c \; p_{true}+\frac{2}{9} \; (p_{false}+p_{straycounts})}{p_{coin}}.
\end{equation}
Here, the parameters have  the same definitions for $p_{true}$,  $p_{false}$, $p_{straycounts}$ and  $p_{coin}$  as in the BBM92 protocol. In the E91 protocol, Alice and Bob randomly choose among three measurement bases. Among the nine possible basis combinations, only two yield correlated outcomes suitable for key generation, while the others are discarded or used for Bell inequality verification. Therefore, the sifting factor is $\tfrac{2}{9}$~\cite{ekert1991quantum} and  QBER will be impacted by a sifting factor of $\frac{2}{9}$.

\section{Keyrate}\label{Keyrate}
Keyrate, in the context of QKD, refers to the rate at which a secure cryptographic key can be generated and shared between two communicating parties, typically referred to as Alice and Bob.  The key rate is measured in bits per pulse and is influenced by factors such as the quality of transmitted quantum states, detection efficiency, channel losses, and potential eavesdropping attempt. The keyrate serves as a benchmark for evaluating the effectiveness and practicality of protocols using QKD. In the key rate calculation, the QBER, denoted by $e$, is taken into account. The QBER reflects the influence of the total transmittance efficiency and incorporates the effects of various loss mechanisms, such as diffraction loss, pointing error and turbulence loss, as well as other factors affecting the security and performance of the QKD system, as discussed above. Accordingly, the key rate provides a comprehensive measure of the effectiveness and efficiency of QKD protocols.

\subsection{\textbf{ Keyrate for BB84}}\label{Keyrate for BB84}
The secure key generation rate against individual attack for the BB84 protocol is given by \cite{lutkenhaus2000security}:
\begin{equation}
    R_{BB84} = \frac{1}{2} p_{click} \{(1-\tau(e_{84})+f(e_{84})(e_{84}\log_2(e_{84})+(1-e_{84})\log_2(1-e_{84}))\}. \label{k_BB84}
\end{equation}
\
$f(e_{84})$ is error correction factor \cite{lutkenhaus2000security},
$\tau$ is fraction of the key to be discarded during privacy amplification,
$\tau(e_{84})$= $\log_2(1+4e_{84}-4e_{84}^2)$ if $e_{84}< 1/2$ and $\tau(e_{84})=1$ if $e_{84}>1/2$.

\subsection{\textbf{ Keyrate for B92}}
The secure key generation rate of the B92 protocol against individual attack can be formulated as \cite{EtenguAbbouWongAbidNortizaSetharaman}:
\begin{equation}
    R_{B92} = \frac{1}{4} p_{click} \{(1-\tau(e_{84})+f(e_{92})(e_{92}\log_2(e_{92})+(1-e_{92})\log_2(1-e_{92}))\}. \label{K_B92}
\end{equation}
In B92, only $25\%$ of the bits transmitted will be detected by Bob, i.e., only $25\%$ of the raw key bits should be kept.
Hence, $\frac{1}{4}$ is the sifting factor. All the other parameters are defined in Sec. \ref{Keyrate for BB84}.
\subsection{\textbf{ Keyrate for BBM92}}
 The keyrate for BBM92 protocol against double blinding attack is given by \cite{PhysRevA.65.052310}:
 \begin{equation}
     R_{BBM92} = \frac{p_{coin}}{2} \{\tau(e_{M92})+f(e_{M92})(e_{M92}\log_2(e_{M92})+(1-e_{M92})\log_2(1-e_{M92}))\}.\label{K_BBM92}
 \end{equation}
 Due to the double-blinding attack discussed in \nameref{Appendix C} , Alice and Bob are unable to detect the presence of Eve, resulting in a complete elimination of information leakage. Within the standard parameter-estimation framework, the information leakage parameter $\tau$ is inferred to be zero under a double-blinding attack, as the observed statistics do not indicate any deviation. However, this inference is not physically valid, since the attack exploits detector behavior that lies outside the assumptions of the underlying security model,
$f(e_{M92})$ is error correction factor,
$p_{coin}$ is the coincidence probability which has already been explained in Sec. \ref{QBER for BBM92}.\\

\subsection{\textbf{ Keyrate for E91}}
The keyrate for E91 protocol against double blinding attack is given by

 \begin{equation}
     R_{E91} = \left( \frac{2}{9} \, p_{\text{coin}} \right)   \{\tau(e_{91})+f(e_{91})(e_{91}\log_2(e_{91})+(1-e_{91})\log_2(1-e_{91}))\}\label{K_E91}.
 \end{equation}
In the E91 protocol, since Alice and Bob each choose among three measurement bases, only two out of the nine possible basis combinations contribute to the raw key, leading to a sifting factor of $\tfrac{2}{9}$
 \cite{ekert1991quantum},  The parameters used in Eq. \eqref{K_E91} have been described  above.

\section{Numerical Results and Discussion }\label{Numerical Results and Discussion}

This section examines the QBER and secret key rate as functions of the zenith angle for the BB84, B92, E91, and BBM92 protocols, considering uplink (night) and downlink (day and night) scenarios. To accurately model atmospheric losses, we employ the circular beam model, which effectively accounts for diffraction losses and pointing errors. In the case of uplink transmission, atmospheric turbulence is also significant, as it primarily affects the initial stage of propagation. Since turbulence is concentrated within the lower 20~km of the atmosphere and is strongest near the Earth's surface, its impact can be neglected for downlink transmission, where it occurs only near the end of the optical path \cite{bourgoin2013comprehensive}. The simulations incorporate the experimental parameters specified in Table \ref{table}.

%~\ref{fig:QKB84}, \ref{fig:QKB92}, \ref{fig:QKM92}, and \ref{fig:QKE91}
\begin{figure}[htbp]
\centering
\subfloat[\label{2a}]{\includegraphics[width=0.52\linewidth]{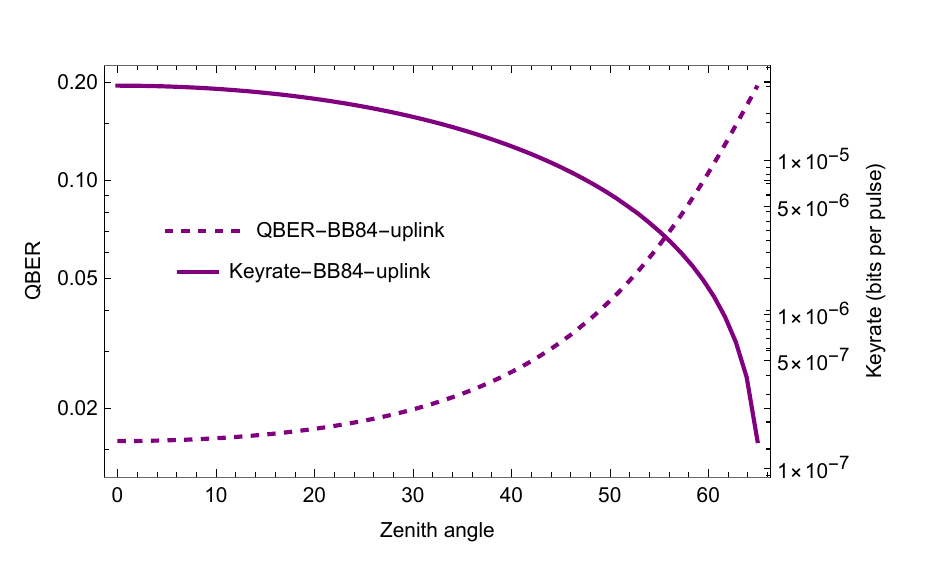}}\\[1ex]
\subfloat[\label{2b}]{\includegraphics[width=0.52\linewidth]{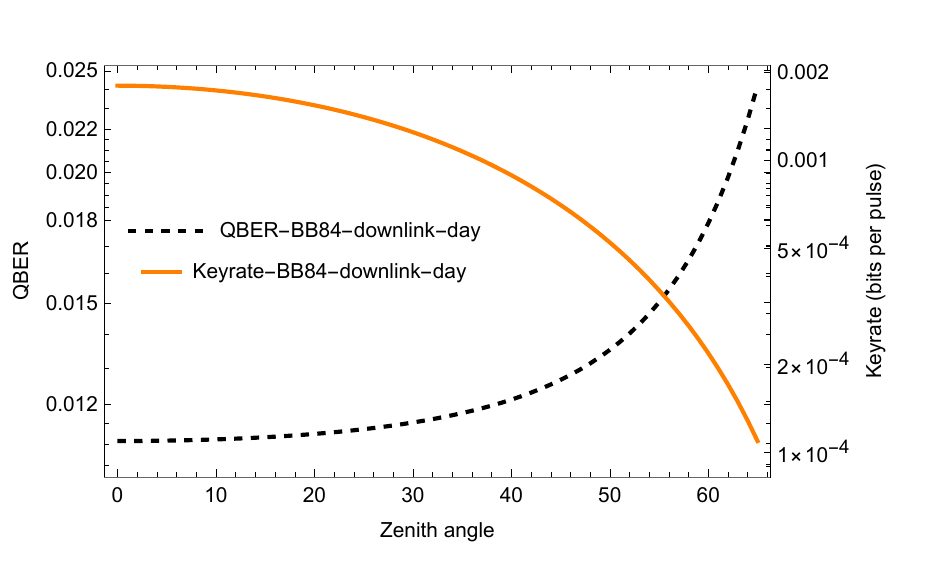}}
\subfloat[\label{2c}]{\includegraphics[width=0.52\linewidth]{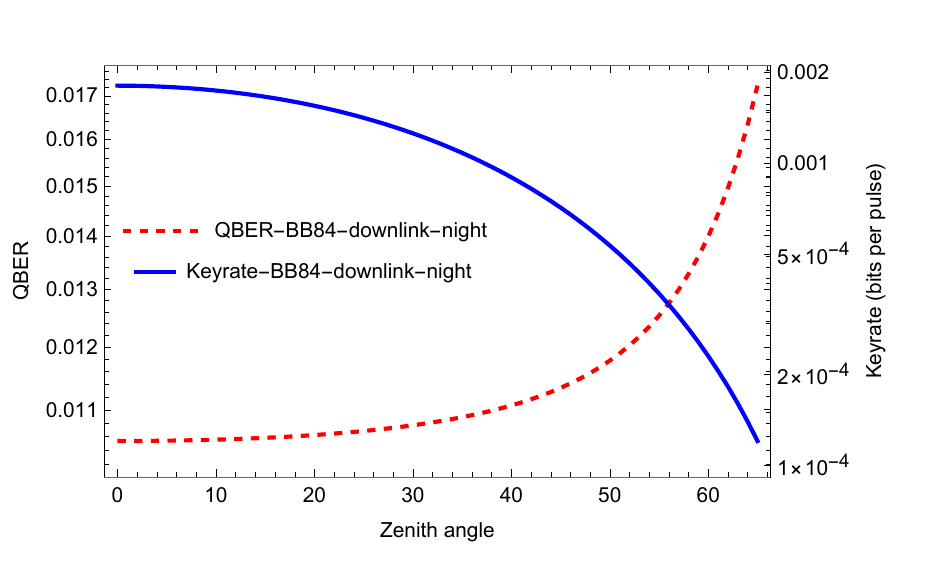}}
%\subfloat[subfig-c's caption]{\includegraphics[height=1.8in]{...}}
\caption{(a) QBER and key rate versus zenith angle for the BB84 protocol in the uplink (night-time). (b) and (c) QBER and key rate versus zenith angle for the BB84 protocol in the downlink (day-time) and downlink (night-time), respectively.}
 
\label{QKB84}
\end{figure}

\begin{figure}[htbp]
\centering
\subfloat[\label{3a}]{\includegraphics[width=0.52\linewidth]{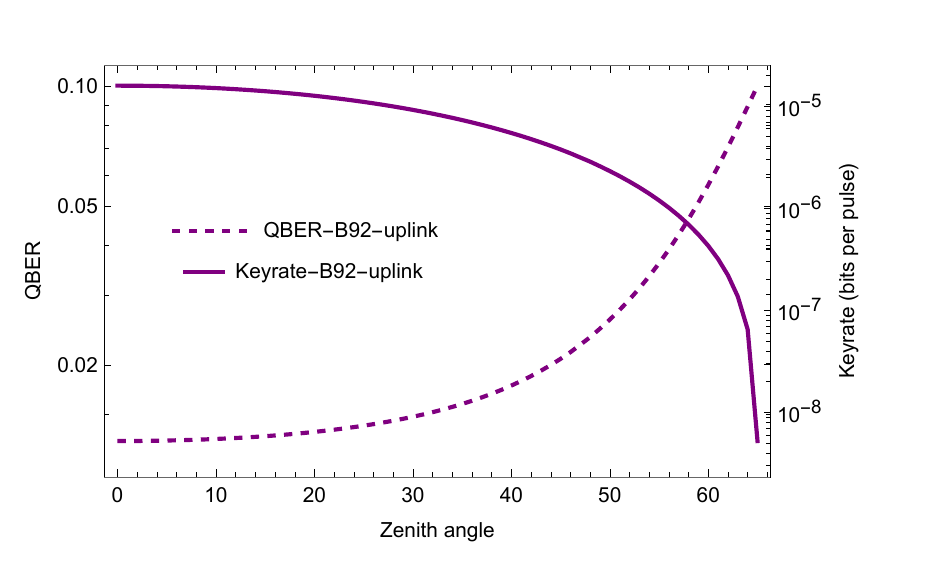}}\\[1ex]
\subfloat[\label{3b}]{\includegraphics[width=0.52\linewidth]{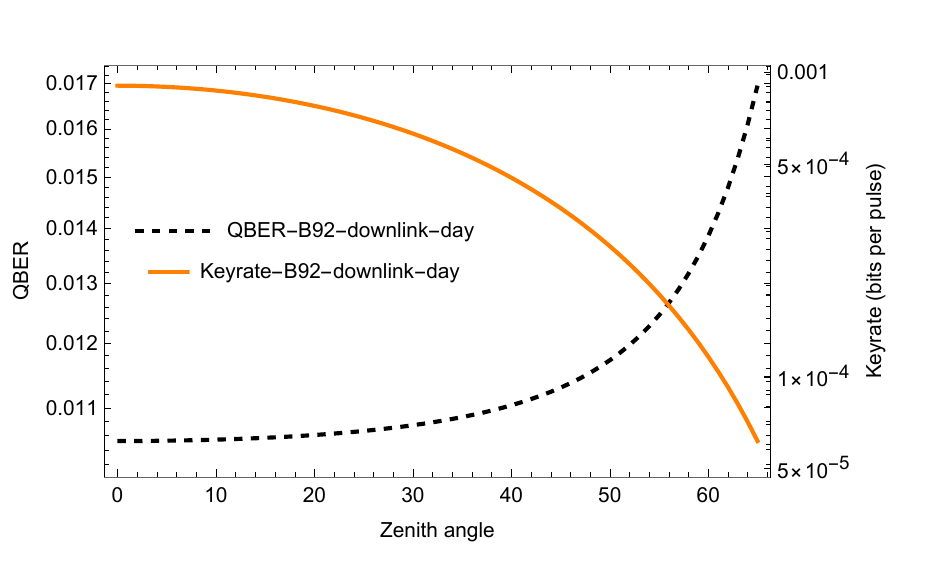}}
\subfloat[\label{3c}]{\includegraphics[width=0.52\linewidth]{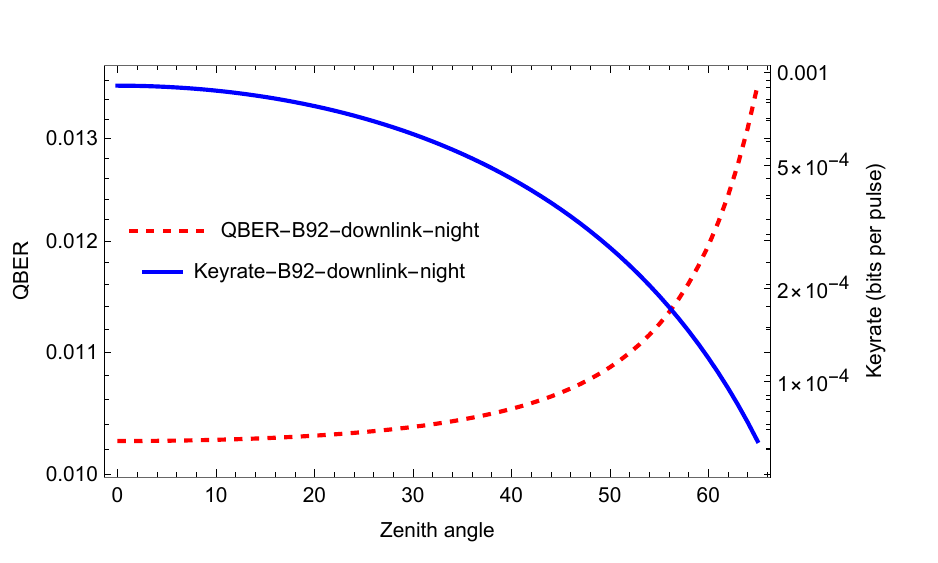}}
%\subfloat[subfig-c's caption]{\includegraphics[height=1.8in]{...}}
\caption{(a) QBER and key rate versus zenith angle for the B92 protocol in the uplink (night-time). (b) and (c) QBER and key rate versus zenith angle for the B92 protocol in the downlink (day-time) and downlink (night-time), respectively.}
\label{QKB92}
\end{figure}

\begin{figure}[htbp]
\centering
\subfloat[\label{4a}]{\includegraphics[width=0.52\linewidth]{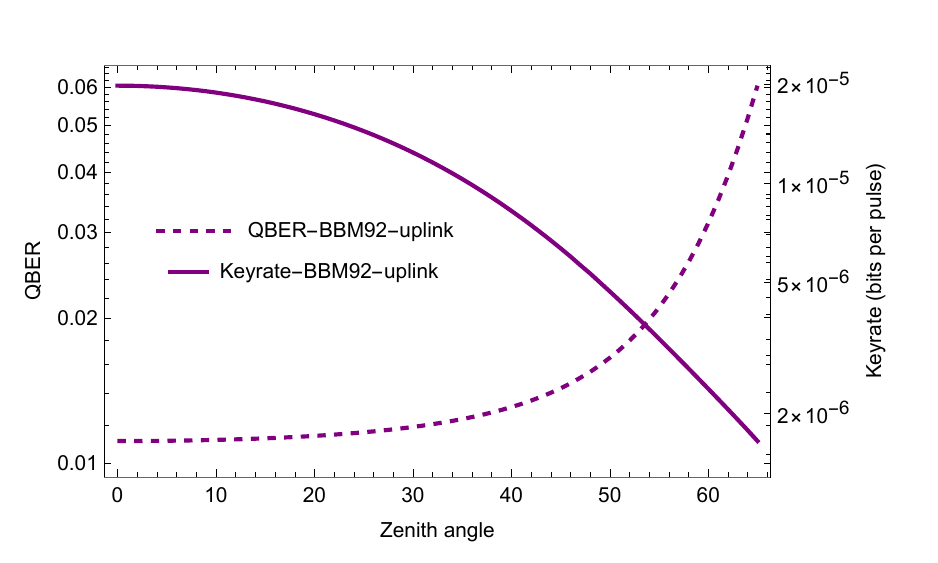}}\\[1ex]
\subfloat[\label{4b}]{\includegraphics[width=0.52\linewidth]{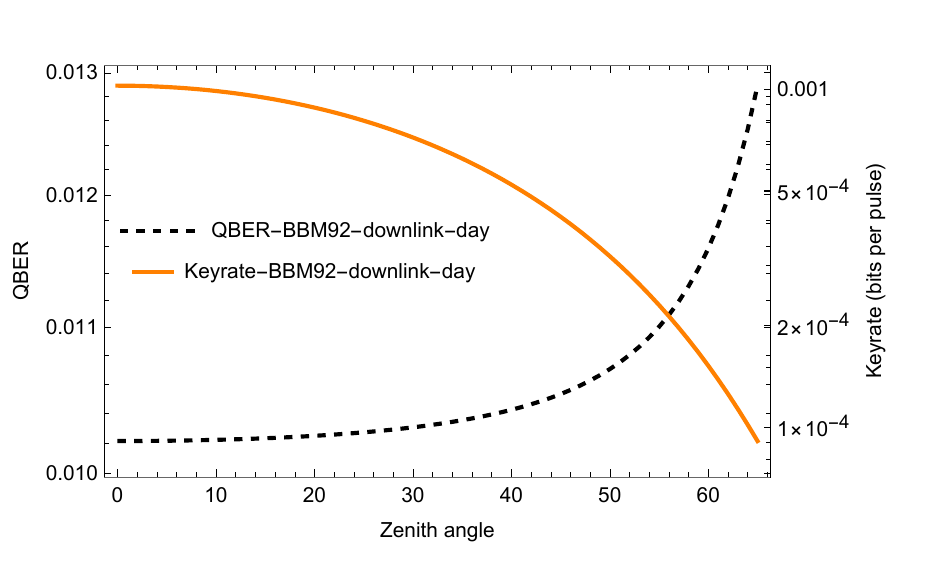}}
\subfloat[\label{4c}]{\includegraphics[width=0.52\linewidth]{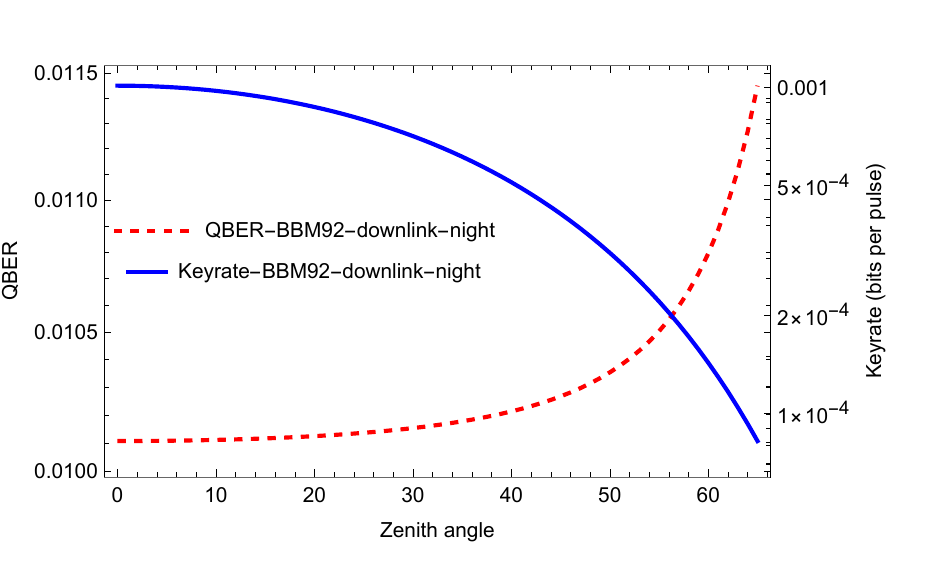}}
%\subfloat[subfig-c's caption]{\includegraphics[height=1.8in]{...}}
\caption{(a) QBER and key rate versus zenith angle for the BBM92 protocol in the uplink (night-time). (b) and (c) QBER and key rate versus zenith angle for the BBM92 protocol in the downlink (day-time) and downlink (night-time), respectively.}
\label{QKM92}
\end{figure}

\begin{figure}[htbp]
\centering
\subfloat[\label{5a}]{\includegraphics[width=0.52\linewidth]{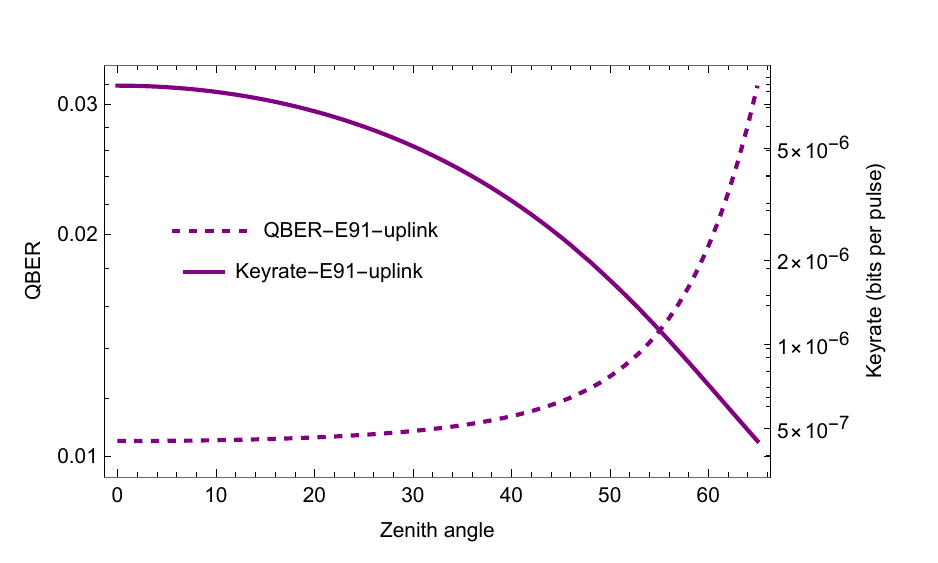}}\\[1ex]
\subfloat[\label{5b}]{\includegraphics[width=0.52\linewidth]{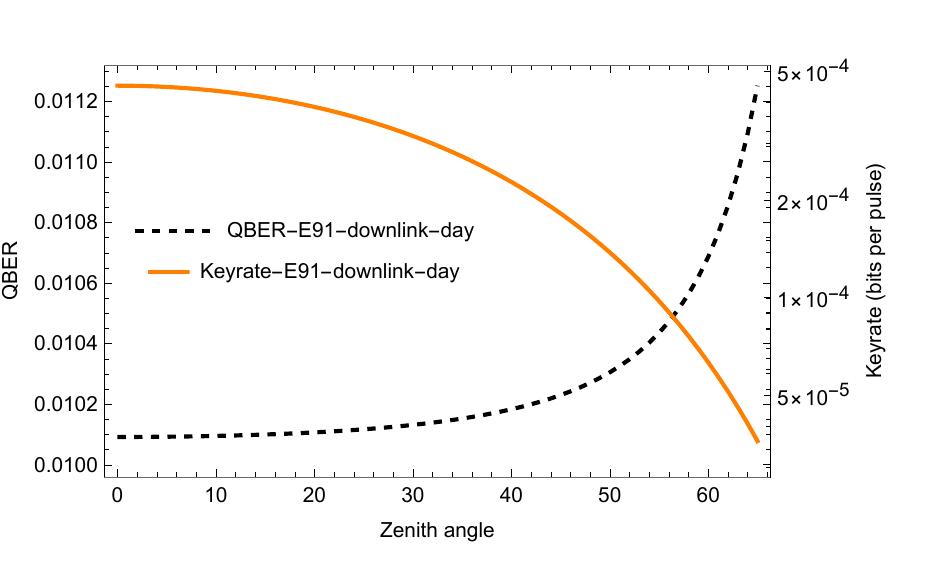}}
\subfloat[\label{5c}]{\includegraphics[width=0.52\linewidth]{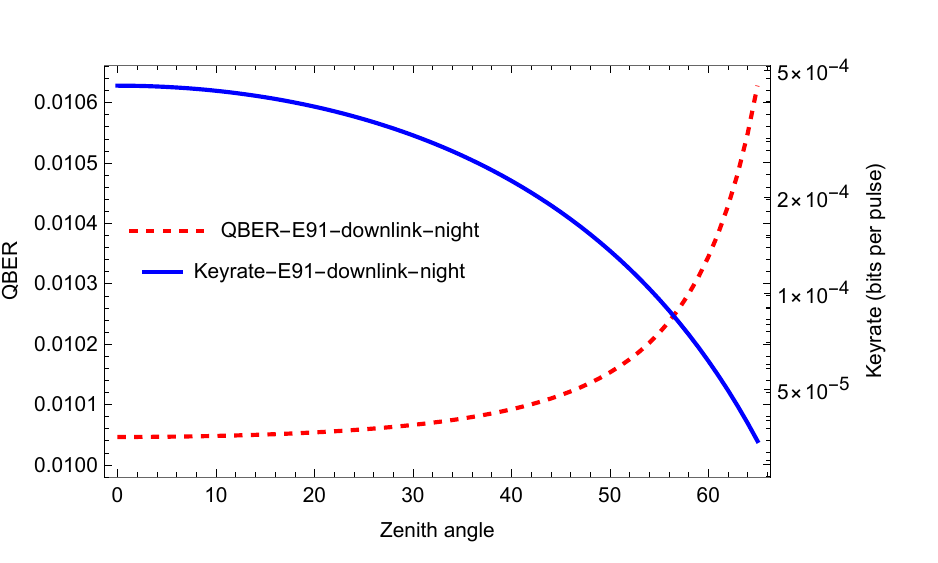}}
%\subfloat[subfig-c's caption]{\includegraphics[height=1.8in]{...}}
\caption{(a) QBER and key rate versus zenith angle for the E91 protocol in the uplink (night-time). (b) and (c) QBER and key rate versus zenith angle for the E91 protocol in the downlink (day-time) and downlink (night-time), respectively.}
\label{QKE91}
\end{figure}
Equations. \eqref{k_BB84} and \eqref{K_B92} give the asymptotic key rate formulas for the BB84 and B92 protocols, while Eqs. \eqref{K_BBM92} and \eqref{K_E91} give the formulas for the entanglement-based protocols BBM92 and E91. These formulas are combined with the circular beam model, which includes the effects of atmospheric transmittance and stray counts for both day and night. This allows us to calculate the key rates for LEO-based quantum communication systems. Figure. \ref{QKB84} shows how the QBER and key rate changes with the zenith angle, taking into account both the transmittance model and stray counts. The analysis is done for uplink (night) and downlink (day and night) cases. In Fig. \ref{2a}, at the zenith position i.e at $0^\circ$ zenith angle, the BB84 uplink exhibits a QBER of $0.0159$ and a key rate of $3.04\times10^{-5}$ bits per pulse. At a higher zenith angle of $55^\circ$, the QBER increases to $0.0631$, while the key rate decreases to $1.30\times10^{-6}$ bits per pulse.  As the zenith angle increases, the signal traverses a longer atmospheric path, leading to greater attenuation due to absorption and scattering. Furthermore, enhanced beam divergence reduces the overlap with the receiving telescope aperture, further lowering the detected photon count. Therefore QBER increases and key rate decreases wih increasing zenith angle.  Figures. \ref{2b} and \ref{2c} depict the QBER and secret-key rate for the BB84 protocol in the downlink day and downlink night configurations, respectively. In Fig. \ref{2b} downlink day, at the zenith position i.e., at $0^\circ$ zenith angle, the QBER is $0.0111$ with a secret-key rate of $1.80\times10^{-3}$ bits per pulse. At a zenith angle of $55^\circ$, the QBER increases to $0.0150$ and the secret-key rate decreases to $3.59\times10^{-4}$ bits per pulse. In Fig. \ref{2c} downlink night, at $0^\circ$ zenith angle, the QBER is $0.0105$ with a secret-key rate of $1.81\times10^{-3}$ bits per pulse. At $55^\circ$, the QBER is $0.0125$, while the secret-key rate is $3.72\times10^{-4}$ bits per pulse.
 In both cases Fig. \ref{2b} and \ref{2c}, the downlink configuration yields a lower QBER and a higher secret-key rate than the uplink scenario shown in Fig. \ref{2a}. When comparing the uplink and downlink scenarios, the uplink experiences beam propagation through the turbulent atmosphere, leading to significant broadening, while the downlink only encounters turbulence during the final stage, resulting in less spreading. As a result, the attenuation in the uplink is more pronounced compared to the downlink due to these differing propagation conditions. Day-time uplink was not considered due to excessive background light \cite{liorni2019satellite}. Furthermore, in the night-time downlink scenario Fig. \ref{2c}, the atmospheric transmittance is reduced due to the higher moisture content compared to daytime conditions. The contrast between day and night operations is significant: during the day, elevated temperatures generate stronger winds and greater mixing between atmospheric layers, which enhances turbulence effects, thereby increasing the QBER and reducing the key rate. In contrast, on clear nights, the lower atmosphere generally exhibits reduced moisture levels relative to daytime, resulting in less beam spreading from scattering particles. Consequently, the night-time scenario achieves a marginally higher key rate and a slightly lower QBER compared to the daytime case.\par

Now, comparing the BB84 and B92 performance for both uplink and downlink configurations, it is observed that BB84 consistently achieves higher secret-key rates while maintaining QBER values that are slightly higher than those of B92. This is because the BB84 protocol employs four quantum states, enabling a larger fraction of detection events to contribute to the sifted key. In contrast, the B92 protocol relies on only two nonorthogonal states, which often lead to inconclusive results. Consequently, its sifting factor is 0.25, lower than that of BB84, which is 0.5. Hence, even with a marginally higher QBER, BB84 retains more sifted bits and tolerates higher error thresholds, resulting in a greater final key rate than B92 under the same channel conditions. In the uplink scenario at the zenith position 0$^\circ$ (Fig.~ \ref{2a}), BB84 records a QBER of about 0.016 with a key rate of approximately $3.04\times10^{-5}$~bits per pulse, whereas B92 (Fig.~\ref{3a}) exhibits slightly lower QBER of around 0.012 and lower key rate of $1.59\times10^{-5}$~bits per pulse, i.e., lower than BB84. For downlink day and night operations, both protocols exhibit lower QBER than in the uplink case; however, BB84 has a clear key rate advantage-for instance, in night-time downlink at 0$^\circ$ (Fig.~\ref{2c}), BB84’s key rate is about $1.81\times10^{-3}$~bits per pulse,  higher than that of B92 (Fig.~\ref{3c}). This performance gap persists at higher zenith angles. For instance, in night-time downlink at 55$^\circ$, BB84 achieves $3.72\times10^{-4}$~bits per pulse versus $1.90\times10^{-4}$~bits per pulse for B92, while in day-time downlink at the same angle, BB84 records $3.59\times10^{-4}$~bits per pulse compared to $1.87\times10^{-4}$~bits per pulse for B92. The sightly lower QBER and higher key rates observed during night-time are primarily due to the absence of background sunlight, which reduces detector noise and photon count errors, as well as the lower atmospheric moisture levels compared to daytime, leading to reduced beam spreading from scattering particles and thereby enhancing both the accuracy of received bits and the overall key generation efficiency. These results highlight BB84’s superior resilience to atmospheric losses and noise, with its four-state encoding enabling more robust and reliable secure key distribution in LEO-based quantum communication across diverse channel conditions.

For the uplink configuration, the BBM92 protocol exhibits QBER values ranging from 0.01 to 0.06 with corresponding secret key rates between $1.64\times10^{-6}$ and $2\times10^{-5}$~bits per pulse (Fig.~\ref{4a}), whereas the E91 protocol records slightly lower QBER values in the range of 0.01-0.03 with key rates between $4.52\times10^{-7}$ and $8.39\times10^{-6}$~bits per pulse (Fig.~\ref{5a}). The relatively higher key rates observed in BBM92 compared to E91 can be attributed to its simpler sifting process and better tolerance to background noise, enabling it to extract more secure bits per detected entangled pair, while the basis correlation requirements in E91 lead to a smaller fraction of usable bits after sifting. In the downlink night-time scenario, BBM92 achieves QBER values in the range of 0.0100-0.0115 with key rates from about $9.09\times10^{-5}$ to $1.03\times10^{-3}$~bits per pulse (Fig.~\ref{4c}), while E91 shows slightly lower QBER values between 0.0100 and 0.0106 with key rates ranging from $3.44\times10^{-5}$  to $4.50\times10^{-4}$~bits per pulse (Fig.~\ref{5c}). In the downlink day-time case, both protocols maintain low QBER values, with BBM92 ranging between 0.010-0.013 and key rates from $9.09\times10^{-5}$ up to $1.03\times10^{-3}$~bits per pulse (Fig.~\ref{4b}), while E91 shows QBER values of 0.0100-0.0112 and key rates in the range of $3.62\times10^{-5}$ to $4.51\times10^{-4}$~bits per pulse (Fig.~\ref{5b}). Overall, E91 generally exhibits marginally lower QBER and lower key rates than BBM92. For both protocols, the key rate decreases with increasing zenith angle because the longer atmospheric slant path increases optical losses and turbulence, reducing the number of entangled photon pairs detected at the receiver.

\section{Conclusion}
This study presented a comparative performance analysis of four QKD protocols—BB84, B92, BBM92, and E91 over LEO satellite-based optical links, considering uplink (night) and downlink (day and night) configurations. By incorporating the circular beam model with diffraction, pointing errors, atmospheric transmittance, and stray counts, we evaluated the QBER and secret-key rate as functions of the zenith angle. The results show that downlink configurations consistently outperform uplink in both QBER and key rate due to reduced turbulence impact. Among the prepare-and-measure protocols, BB84 achieves higher key rates than B92 owing to its four-state encoding, which yields a larger sifted key fraction despite slightly higher QBER. Similarly, in entanglement-based schemes, BBM92 attains higher key rates than E91 due to its simpler sifting process and better tolerance to background noise, though E91 exhibits marginally lower QBER. Across all protocols, key rates decrease with increasing zenith angle as a result of longer atmospheric paths and greater optical losses. Night-time operations yield better performance than daytime due to reduced background light and detector noise. These findings highlight BB84 and BBM92 as more suitable choices for high-rate, long-distance satellite QKD, while also providing insights into the trade-offs between protocol design, link configuration, and atmospheric conditions in space-based quantum communication.

\section{Limitations and Future Enhancements}

In the present study, we have analyzed both prepare-and-measure and entanglement-based QKD protocols using a Gaussian beam framework, where the beam is approximated as circular at the receiver plane, to evaluate the secure key rate. Although this approach provides a realistic and widely accepted representation of optical beam propagation in satellite-based QKD links, but it does not account for other QKD protocols or more refined beam propagation models. In particular, alternative beam approximations, such as the elliptical beam model, which more accurately captures the effects of atmospheric turbulence and beam deformation, have not been considered in this work. Future work can be extended to incorporate such models, which could provide deeper insights into channel behavior and system performance under varying atmospheric conditions. Furthermore, the present analysis is limited to a selected set of discrete-variable QKD protocols. Future work could extend this study to include a broader range of protocols, such as continuous-variable (CV) QKD schemes, enabling a more comprehensive comparative assessment across different implementations. Additionally, the current framework focuses primarily on average system performance metrics. A more detailed statistical analysis, including the probability distribution of the secure key rate, could provide further understanding of performance fluctuations and reliability under realistic channel conditions. Finally, this work can be extended to more practical deployment scenarios, such as CubeSat-based QKD systems, where additional constraints related to size, weight, power, and pointing accuracy may significantly impact system performance.

\subsection*{Acknowledgements}
The author would like to thank CSIR for the fellowship support. SB acknowledges support from Interdisciplinary Cyber Physical Systems (ICPS) programme of the Department of Science and Technology (DST), India, Grant No.:DST/ICPS/QuST/Theme-1/2019/6. SB also acknowledges the valuable contribution of the Defense Research and Development Organization (DRDO). The author also extends thanks to Arindam Dutta, Prof. V. Narayanan, Vedant and Prof. Anirban Pathak for valuable suggestions and insightful discussions.

\subsubsection*{Declarations}
The authors declare no conflicts of interest related to this research.

\bibliographystyle{unsrt}
\bibliography{source}

\appendix

\section*{Appendix A} \label{Appendix C}
\subsection*{Blinding Attacks (Single and Double Blinding Attacks)}

In the context of the BBM92 protocol, the existing blinding attack are of the intercept and resend type. In this type of attack, a malicious entity, often referred to as Eve, intercepts the signal that was originally intended for Bob. Eve then proceeds to perform measurements using random bases in order to obtain the raw key, just as Bob would have done in the intended communication process.
To conceal her presence, Eve forwards a signal to Bob whenever she successfully obtains a measurement result. This signal ensures that Bob receives an identical outcome, while in the case of diagonal alignment, no detection occurs at all.
 In practical implementation using QKD devices \cite{Adenier2011DoubleBO}, Eve employs techniques to blind Bob's detectors to single-photon detection. She achieves this by manipulating the detectors to shift from Geiger mode to linear mode, where a detector only registers a click if the incoming signal intensity exceeds a preset discriminator threshold, denoted as $I_{th}$. After each detection, Eve sends a bright pulse with linear polarization aligned to her own measurement result. When Eve and Bob randomly select identical measurement bases, the pulse deterministically generates a click in one of Bob's detectors. This ensures that Bob's measurement outcomes match those of Eve because the pulse is either fully reflected or transmitted at Bob's polarizing beamsplitter. However, to prevent double counting and incorrect results when Eve and Bob randomly select bases that are diagonal to each other, Eve adjusts the intensity of the pulses to be lower than twice the threshold intensity of the detectors. Consequently, the pulse is split in half at Bob's polarizing beamsplitter, resulting in an output that is insufficient to surpass the threshold and produce a click in either of Bob's detectors.
The objective of the attack is for Eve to obtain an exact replica of Bob's key at the conclusion of the raw key distribution process. If Alice and Bob are sufficiently satisfied with the measured QBER on a subset of the key, Eve can eavesdrop on the error correction protocol that Alice and Bob employ. By performing the same operations as Bob during the error correction phase, Eve can successfully acquire an exact copy of the sifted key in the end.
One limitation of single-blinding attacks is that, on average, Bob's resulting key size is reduced by half compared to what he would have obtained without the attack. This reduction occurs because approximately half of the time, the randomly chosen measurement bases of Eve and Bob turn out to be diagonally opposite to each other. Consequently, Bob's detectors do not register any clicks in such cases. Therefore, the efficiency of this attack, by design, is fundamentally limited to $50\%$ on Bob's side.\par
Here the proposed double-blinding attack involves a similar implementation to the single-blinding attack, but with the key difference that Eve blinds all detectors on both sides instead of just Bob's detectors. Due to the double-blinding attack, Alice and Bob are unable to detect the presence of Eve, resulting in a complete elimination of information leakage. In other words, the measure of information leakage, denoted as  $\tau$ becomes zero in this scenario.

\end{document}